&latex209
\documentstyle[multicol,overcite,pre,aps,epsf]{revtex}
\newcommand{\wdtxt}{\end{multicols}\widetext\hspace{-\parindent}\hrulefill 
\hspace{2.0in}\mbox{}}
\newcommand{\nrtxt}{\mbox{}\hspace{2.0in}\hrulefill\mbox{}\begin{multicols}{2}
\narrowtext\hspace{-\parindent}}
\begin{document}		    
\draft

\title{Fluctuation Effects on Quadratic Autocatalysis Fronts}

\author{Mikhail V. Velikanov and Raymond Kapral\\ 
Chemical Physics Theory Group, Department of Chemistry,\\
University of Toronto, Toronto, Ontario M5S 3H6, Canada.}

\maketitle

\begin{abstract}
	A Markov chain model for spatially distributed autocatalytic systems
with a quadratic reaction rate is considered. An approximate solution for the
local probability distribution is obtained in the form of a perturbation 
expansion for the regimes where diffusion is relatively fast. Using this 
approximate distribution, properties of the chemical wave fronts found in 
these autocatalytic systems are studied, and deviations of the minimum propagation
velocity and the concentration profile from deterministic predictions are 
analyzed. A comparison with numerical results from lattice-gas automaton 
simulations is also provided.
\end{abstract}


\begin{multicols}{2}		  
\narrowtext

\section{Introduction}
\label{intro}
	Chemical wave fronts are simple examples of spatiotemporal patterns 
supported by nonequilibrium chemical systems \cite{fife,field,baras,mikhailov}. 
The dynamics of such fronts is generally described by a set of deterministic 
reaction-diffusion equations; a wave front is defined as a special solution of 
these equations, unformly translating in space with constant velocity and 
connecting two distinct homogeneous stationary states. In particular, the wave 
front solutions of the reaction-diffusion equation,
\begin{equation}							 
\label{FK_eq}
{\partial c \over \partial t} = \nabla^2 c + c \: \left( 1 - c \right)\;,
\end{equation}
have been studied extensively. Here $c$ is the (scalar) order parameter field 
($c$ varies between $0$ and $1$), and $t$ is the time variable. Equation 
(\ref{FK_eq}) was first considered by Fisher \cite{fisher} and Kolmogorov 
{\it et al.} \cite{kolmogorov} in the context of population dynamics and has 
since found applications in a variety of other disciplines, including nonlinear 
chemical kinetics \cite{droz,scott,murray,zeldovich}. From a chemical viewpoint, 
equations of this type describe the evolution of a spatially distributed 
autocatalytic system with a quadratic reaction rate (referred to in this paper 
as a quadratic autocatalysis system). 

	It is well known that, given appropriate initial conditions, eq. 
(\ref{FK_eq}) admits (one-dimensional) wave front solutions of the form 
$c = c(x - v t)$, where $v$ is the front velocity and $x$ is the spatial 
coordinate in the direction of front movement. The front velocity must not be 
less than the minimum velocity $v_{{\rm min}} = 2$ in order for the wave front 
to exist \cite{fife,murray}. Further analysis (see ref. \cite{billingham} for 
details) shows that the solution with $v = v_{{\rm min}}$ is marginally stable 
and, therefore, any wave front initiated with a sufficiently steep initial 
profile will eventually relax to this minimum-velocity solution. No exact 
analytic solution of eq. (\ref{FK_eq}) is known for general $v$; however, it 
is possible to derive the following approximate form for the front profile, 
which is uniformly valid for $v \geq v_{{\rm min}}$ \cite{murray}:
\wdtxt
\begin{equation}
\label{app_profile}
c(z) = {1 \over 1 + e^{z/v}} + {e^{z/v} \over v^2 \: \left( 1 + e^{z/v} 
\right)^2} \: \ln \left[ {4 \: e^{z/v} \over \left( 1 + e^{z/v} \right)^2} 
\right],
\end{equation}
\nrtxt
where $z = x - v t$ is the spatial coordinate of the reference frame moving 
with the wave front.

	This deterministic description, although capable of capturing the gross 
features of the front dynamics, is less successful with respect to finer 
details. For instance, reaction-diffusion equations do not provide an adequate 
framework for understanding the influence of fluctuations intrinsic to the 
local reactive and diffusive dynamics on the macroscopic properties of the 
wave front. This is due to the fact that reaction-diffusion equations are 
essentially mean-field equations in which all correlations built by the local 
fluctuating dynamics are neglected. However, recent numerical studies on 
chemical fronts have shown that fluctuations significantly affect both the 
propagation velocity and the concentration profile of the wave front 
\cite{lemarchand,lesne,karzazi,gorecki,sanders,derrida}.

	In this paper we present a theory which systematically accounts for 
the effects of fluctuations on the properties of the wave fronts in quadratic 
autocatalysis systems. Our analysis is based on the perturbation theoretic
formalism developed earlier and applied to an oscillatory reaction-diffusion
system \cite{theory}. The system is described by a Markov 
chain model \cite{cox,gardiner} whose mean-field kinetic equation is the 
generalized Fisher-Kolmogorov equation. We use a Chapman-Enskog-like perturbative 
technique \cite{uhlenbeck} to extract the corrections to the mean-field local 
probability distribution due to fluctuations in the local dynamics. The formal 
expansion parameter $\gamma$ gauges the ratio of characteristic time scales of 
diffusion and reaction processes, and is small when diffusion is fast enough, 
i.e. the deviation from the mean-field behaviour is small.  
The kinetic equation for the local concentration obtained from the generalized
local distribution contains a reaction rate term which is not quadratic in the 
concentration. We further analyze the wave front solutions of the generalized 
kinetic equation and determine the dependence of the front properties on the 
diffusion coefficient. We also compare these results with the deterministic 
predictions and the data obtained in lattice-gas automaton simulations of the 
quadratic autocatalysis system.

	The paper is organized in the following manner. In Sec.~\ref{model} we 
introduce the Markov chain model for quadratic autocatalysis systems and review
the mean-field results for the properties of the wave fronts in this model. In
Sec. \ref{results}, using the perturbative technique developed in Ref.~\cite{theory},
the wave fronts are analyzed in a framework that goes beyond mean-field theory. 
The dependence of the minimum propagation velocity and the front shape on the 
diffusion coefficient, including the effects of fluctuations, is derived and 
compared to the mean-field results. Section \ref{LGA} presents the results of 
lattice-gas automaton simulations of a quadratic autocatalysis system
and compares these results with the predictions of the perturbation theory. Finally, Sec. 
\ref{conclusions} contains a discussion of the results of our study.

\section{Markov Chain Model}
\label{model}
	We consider a system which consists of a chemical species, $A$, diffusing 
and reacting in solution. We assume a discrete-space, discrete-time description 
of the system in which space is partitioned into cells of fixed volume and 
time into intervals of unit length. The cells can be alternatively viewed 
as nodes of a lattice; reactant particles residing at a node may react according 
to the reaction mechanism, with probability determined by combinatorial 
rules, or execute diffusive jumps to the neighboring nodes. Furthermore, we 
impose exclusion principle, according to which a node cannot be occupied by 
more than $N$ reactant particles. 

	The following ``particle-hole'' reaction mechanism is used in this 
model:
\begin{eqnarray}
A + A^* &\mathrel{\mathop{\kern0pt {\rightarrow}}\limits^{k_1}_{}}& 
2 A, \nonumber \\
A &\mathrel{\mathop{\kern0pt {\rightarrow}}\limits^{k_2}_{}}&
\emptyset. \nonumber
\end{eqnarray}
Here $k_1$ and $k_2$ are rate coefficients and an asterisk denotes a 
vacancy, i.e. an empty space at a node that can be occupied by a particle of 
$A$. The concentrations of pool chemical species (i.e. species whose concentrations 
are fixed by constraints) are incorporated in the rate constants. Reactions of 
this type that depend on the concentrations of both particles and particle 
vacancies are common in surface chemistry and in biochemistry. The solvent 
particles are assumed to be chemically inert. At any time the state of the 
system is completely specified by the number of the particles of $A$ occupying 
every node. In the following, we set $k_1 = 1$; one can easily show that this 
is always possible by appropriate rescaling of time. The probabilities per 
unit time of the two reactions, determined by the combinatorics, are, 
respectively,
$$
p_1={h \over N-1} \: a({\bf r}) \: \left( N - a({\bf r}) \right) \: , \;\;
p_2=h k_2 \: a({\bf r}) \: .
$$ 
Here $h$ is a parameter that sets the time scale of the Markov chain, and 
$a({\bf r})$ is the total number of particles at the node with position vector 
${\bf r}$. The factor $(N-1)^{-1}$ is included so that the mean-field kinetic 
equation (i.e. reaction-diffusion equation) for the concentration of $A$ can 
be written in a neat analytic form (see discussion below). 

	The inert solvent particles serve to randomize of the diffusive jumps 
that the reactant particles execute. The probability that a particle at the 
node with position vector ${\bf r}$ will execute a diffusive jump within a unit 
time interval is equal to the ratio $a({\bf r})/N$. The direction of the jumps 
is selected at random for each time interval and is the same for all nodes in 
the system. One can show that this diffusion mechanism yields the correct form 
of the diffusion equation for the local concentration. 

	The Markov chain which describes evolution of the full probability 
distribution function is constructed from the successive application of the 
reaction and diffusion Markov chains. The numerical simulations reported in 
Sec. \ref{LGA} utilize this full Markov chain, without any further assumption 
or simplification. 

\subsection{Local Markov chain dynamics}
	To facilitate theoretical analysis of the Markov chain model an 
additional assumption is employed. Numerical studies of the diffusion Markov 
chain \cite{theory,malevan} showed that for systems with spatial dimensions
two or greater the diffusion rule described above leads to rapid de-correlation 
of the particle number densities at different spatial locations. Consequently,
the full probability distribution for the entire system may be factored into 
a product of single-node probability distributions. In the following, we 
assume that the system's dimensionality is large enough so that this factorization
approximation is valid and, for sufficiently long times, the de-correlation 
allows description of the evolution in terms of the {\it local} probability 
distribution function $P(a({\bf r}),n)$. 

	The evolution is due to two competing processes, namely local reactive 
events and diffusive transport of particles. Correspondingly, the evolution 
equation for the local distribution function $P(a({\bf r}),n)$ can be written 
in the following form:
\eject
\wdtxt
\begin{equation}
\label{RD_markov}			
P(a({\bf r}),n+1) - P(a({\bf r}),n) = \left( \gamma \hat{W}^R + \hat{W}^D 
\right) P(a({\bf r}),n).
\end{equation}
\nrtxt
Here $n$ is the (integer) time variable and $\hat{W}^D$ and $\hat{W}^R$ are 
the evolution operators for pure diffusion and reaction processes, respectively.
The dimensionless parameter $\gamma$ gauges the relative contributions of these 
two processes to the overall dynamics. A natural choice for the definition of 
$\gamma$ is the ratio of characteristic time scales of the diffusion and 
reaction processes.

	The matrix elements of the diffusion evolution operator $\hat{W}^D$ 
are \cite{malevan}
\wdtxt
\begin{eqnarray}
\label{Wdiff}
W^D_{a({\bf r}),a'({\bf r})} &=& {\chi({\bf r},n) \over N}
\left(1 - {a'({\bf r}) \over N} \right) \delta_{a'({\bf r}),a({\bf r}) - 1}
+ {a'({\bf r}) \over N} \left(1 - {\chi({\bf r},n) \over N}\right)
\delta_{a'({\bf r}),a({\bf r}) + 1} \nonumber \\
&-& \left[ {\chi({\bf r},n) \over N} \left(1 - {a'({\bf r}) \over N} 
\right) + {a'({\bf r}) \over N} \left(1 - {\chi({\bf r},n) \over N} 
\right) \right] \delta_{a'({\bf r}),a({\bf r})}.
\end{eqnarray}
\nrtxt						       
Here $\chi({\bf r},n)$ is the mean particle number density (i.e. concentration)
averaged over the immediate neighbourhood of node ${\bf r}$, i.e.
\begin{equation}
\chi({\bf r},n) = {1 \over m} \sum_{{\bf r}' \in \cal N({\bf r})}^{} 
\overline{a}({\bf r}',n), 
\end{equation}
where $m$ is the coordination number of the lattice, $\cal N({\bf r})$ is the
neighbourhood of node ${\bf r}$ and $\overline{a}({\bf r}',n)$ is concentration 
at node ${\bf r}'$ at time $n$, 
$$
\overline{a}({\bf r}',n) = \sum_{a({\bf r}')} a({\bf r}') P(a({\bf r}'), n) \; .
$$
One can show that the stationary local distribution for the $\hat{W}^D$ 
operator alone is binomial,
\wdtxt
\begin{equation}
\label{binom}		
P^D_s(a({\bf r})) = {N \choose a({\bf r})} \left({\overline{a}({\bf r}) \over 
N} \right)^{a({\bf r})} \left(1 - {\overline{a}({\bf r}) \over N} \right)^{N 
- a({\bf r})}.
\end{equation}
\nrtxt
\hspace{\parindent}
The matrix elements of the reaction evolution operator $\hat{W}^R$ are
\wdtxt
\begin{eqnarray}
\label{WReact}
W^R_{a({\bf r}),a'({\bf r})} &=& {h \over N-1} \: a'({\bf r}) \Bigl( N -
a'({\bf r}) \Bigr) \: \delta_{a'({\bf r}),a({\bf r}) - 1} +  h k_2 \: 
a'({\bf r}) \: \delta_{a'({\bf r}),a({\bf r}) + 1} \nonumber \\
&-& \left[ {h \over N-1} \: a'({\bf r}) \Bigl( N - a'({\bf r}) \Bigr) +
h k_2 \: a'({\bf r}) \right] \: \delta_{a'({\bf r}),a({\bf r})}.
\end{eqnarray}						
\nrtxt
Note that the operator $\hat{W}^R$ satisfies the exclusion principle by 
construction since the probability of creating a particle of $A$ at a fully 
occupied node vanishes. 

	If $k_2 > 0$, the stationary distribution for the pure reaction Markov 
chain is $\delta_{a({\bf r}),0}$. This strongly correlated distribution is the 
consequence of the absorbing boundary at $a({\bf r}) = 0$ inherent in the pure 
reaction dynamics. Indeed, as one can see from (\ref{WReact}), once a system 
evolving under the pure reaction Markov chain reaches the state with 
$a({\bf r}) = 0$ (i.e. totally empty), it remains trapped in that state for all 
future times. In the special case $k_2 = 0$, there is an additional absorbing 
boundary at $a({\bf r}) = N$, hence the stationary distribution has the form 
$n \delta_{a({\bf r}),0} + l \delta_{a({\bf r}),N}$, where $n$ and $l$ are 
constants satisfying $n+l=1$ and determined by the initial conditions.

	In a spatially-distributed system with finite diffusion coefficient,
the analytic form of the local probability distribution is determined by the 
interplay between the reactive and diffusive processes described by the 
operators $\hat{W}^R$ and $\hat{W}^D$. Namely, as the diffusion coefficient
increases, the local distribution tends to the uncorrelated binomial form
prescribed by the pure diffusion process, with time dependence incorporated
trivially through the mean particle density, {\it viz}
\wdtxt
\begin{equation}
\label{binomial}		
P_B(a({\bf r}),n) = {N \choose a({\bf r})} \left({\overline{a}({\bf r}, n) \over 
N} \right)^{a({\bf r})} \left(1 - {\overline{a}({\bf r}, n) \over N} \right)^{N 
- a({\bf r})}.
\end{equation}
\nrtxt
Conversely, as the diffusion coefficient tends to zero, correlations characteristic
of the stationary distribution of the pure reaction Markov chain contaminate
the binomial form (\ref{binomial}) to an increasing degree. The subject of this
paper is the effect of these correlations on the properties of the macroscopic
structures, such as wave fronts, existing in the system.

\subsection{Mean-field dynamics}
	Consider the mean-field dynamics of the reaction-diffusion Markov chain. 
The mean-field local distribution function is the time-dependent binomial 
(\ref{binomial}); the kinetic equation for concentration of $A$ can be easily 
obtained using it and eq. (\ref{RD_markov}); after some analysis, we find
\wdtxt
\begin{eqnarray}
\label{mf_RD}
\overline{a}({\bf r}, n+1) - \overline{a}({\bf r}, n) &=& \sum_{a = 0}^{N} 
a({\bf r}) \: \left( \hat{W}^D + \gamma \hat{W}^R \right) P_B(a({\bf r}),n) 
\nonumber \\
&=& {1 \over m N} \: \Delta \overline{a}({\bf r},n) + \gamma h \: 
\overline{a}({\bf r}, n) \: \left( \beta - {\overline{a}({\bf r}, n) \over N} 
\right) \\
&\equiv& \tilde{D} \: \Delta \overline{a}({\bf r},n) + \gamma R(\overline{a}({\bf r}, n)). 
\nonumber
\end{eqnarray}
\nrtxt

	Here $\beta = 1 - k_2$, $\tilde{D} = (m N)^{-1}$, $\Delta$ is the 
discrete Laplacian operator and $R(\overline{a}({\bf r}, n))$ denotes the 
mass-action law terms. Introducing the rescaled variables, 
$$
h \rightarrow \gamma \beta h, \;\;\; \overline{a}({\bf r}, n) \rightarrow 
{\overline{a}({\bf r}, n) \over N \beta} \: ,
$$ 
and the length unit, $L = \sqrt{h / \tilde{D}}$, one can rewrite eq.~(\ref{mf_RD}) 
as follows:
\wdtxt
\begin{equation}
\label{mf_RD_resc}
\overline{a}({\bf r}, n+1) - \overline{a}({\bf r}, n) = {h \over L^2} \: 
\Delta \overline{a}({\bf r}, n) + h \: \overline{a}({\bf r}, n) \: \Bigl(
1 - \overline{a}({\bf r}, n) \Bigr).
\end{equation}
\nrtxt
Clearly, eq. (\ref{mf_RD_resc}) is just a discretized form of eq. (\ref{FK_eq}).
\vspace{-8 pt}
\parbox[b]{8 cm}{
\section{Perturbation Theory for the Wave Front Solutions}
\label{results}
\vspace{-10 pt}
	We now turn our attention to the analysis of the evolution equation 
(\ref{RD_markov}). We derive an approximate kinetic equation for the local 
concentration field valid for small $\gamma$ and analyze the properties of 
its wave front solutions.
\subsection{Local concentration dynamics for $\gamma \ll 1$}
\hspace{\parindent}
We seek the solution of eq. (\ref{RD_markov}) in the form of a regular
perturbation series with time-dependent binomial distribution as the leading
term,}
\vspace*{30 pt}
\begin{equation}
\label{p_series}
P(a({\bf r}),n) = P_B(a({\bf r}),n) + \sum_{k=1}^{\infty} \gamma^k \: 
P_k(a({\bf r}),n).
\end{equation}
In the following, we will keep only the leading and first-order terms of 
the series (\ref{p_series}). The starting point of our analysis is the 
evolution equation for the first-order term, $P_1(a({\bf r}),n)$, which is 
obtained by substituting eq. (\ref{p_series}) into eq. (\ref{RD_markov}),
collecting terms of similar order in $\gamma$ and using the solvability 
conditions (cf. eq. $(20)$ of Ref.~\cite{theory} and discussion therein), 
{\it viz}
\vspace*{18 pt}
\wdtxt
\begin{equation}
\label{p1_ev}
P_1(a({\bf r}),n+1) - P_1(a({\bf r}),n) = \hat{W}^D P_1(a({\bf r}),n) + 
\hat{S} P_B(a({\bf r}),n),
\end{equation}
\nrtxt
where the operator $\hat{S}$ is defined as follows:
\begin{equation}
\label{S_def}
\hat{S} = \hat{W}^R -  R(\overline{a}({\bf r},n)) \: {\partial \over \partial 
\overline{a}} \;\; .
\end{equation}

	Suppose that $P_1(a({\bf r}),n)$ changes slowly with time. Then, we 
can write the formal solution of eq. (\ref{p1_ev}) as
\begin{equation}						    
\label{p1}
P_1(a({\bf r}),n) = - \: \left( \hat{W}^{D} \right)^{-1} \: \hat{S} P_B(a({\bf r}),n),
\end{equation}							    
where $\left( \hat{W}^{D} \right)^{-1}$ is the operator reciprocal to $\hat{W}^D$. 
Using (\ref{p1}) and (\ref{RD_markov}), one can now determine the dynamics of 
the local concentration and we obtain
\wdtxt
\begin{eqnarray}		
\label{a_bar_ev}
\overline{a}({\bf r},n+1) - \overline{a}({\bf r},n) &=& \sum_{a({\bf r}) = 0}^{N}
a({\bf r}) \: \left( \gamma \hat{W}^R + \hat{W}^D \right) \: \Bigl( 
P_B(a({\bf r}),n) + \gamma P_1(a({\bf r}),n) \Bigr) \nonumber \\
&=& {1 \over m N} \: \Delta \overline{a}({\bf r},n) + \gamma R(\overline{a}(
{\bf r})) - \gamma^2 \sum_{a({\bf r}) = 0}^{N} a({\bf r}) \: \hat{W}^R 
\left( \hat{W}^{D} \right)^{-1} \hat{S} P_B(a({\bf r}),n) \: .
\end{eqnarray}
\nrtxt

	The calculation of the $\cal{O}$$(\gamma^2)$ correction term in eq. 
(\ref{a_bar_ev}) is presented in the Appendix. We find
\wdtxt
\begin{equation}
\label{corr_term}    
\sum_{a({\bf r}) = 0}^{N} a({\bf r}) \: \hat{W}^R \left( \hat{W}^{D} \right)^{-1} 
\hat{S} P_B(a({\bf r}),n) = {N h^2 \over N-1} \; \overline{a}({\bf r},n) \left(
1 - {\overline{a}({\bf r},n) \over N} \right)^2 \; .
\end{equation}
\nrtxt
The final form of the kinetic equation for $\overline{a}({\bf r},n)$ is 
obtained by substituting (\ref{corr_term}) into (\ref{a_bar_ev}) and transforming 
the concentration variable, $\overline{a}({\bf r},n) \rightarrow \overline{a}({\bf r},n)/N$. 
Moreover, at this point we can set the ordering parameter $\gamma$ to
unity. In this way we come to
\wdtxt
\begin{equation}
\label{a_bar_ev_final}
\overline{a}({\bf r},n+1) - \overline{a}({\bf r},n) = \tilde{D} \: \Delta 
\overline{a}({\bf r},n) \: + \: h \; \overline{a}({\bf r},n) \: \Bigl( \beta 
- \overline{a}({\bf r},n) \Bigr) \: - \: {N h^2 \over N-1} \; \overline{a}
({\bf r},n) \: \Bigl( 1 - \overline{a}({\bf r},n) \Bigr)^2 \: ,
\end{equation}
\nrtxt
where $\beta = 1 - k_2$. Note that eq. (\ref{a_bar_ev_final}) can be derived 
from the exact solution of (\ref{p1_ev}) by an asymptotic expansion for small 
$\gamma$ of the integral terms in the corresponding kinetic equation (cf. eq. $(25)$ 
of Ref.~\cite{theory} and discussion that follows).

	For any fixed $h$, eq. (\ref{a_bar_ev_final}) can be viewed as a 
discretized version of the following partial differential equation (written 
here for the case of a planar front):
\wdtxt
\begin{equation}
\label{a_bar_ev_ctime}
{\partial \overline{a}(x,t) \over \partial t} = D \: {\partial^2 \overline{a}(x,t)
\over \partial x^2} + \overline{a}(x,t) \: \Bigl( \beta - \overline{a}(x,t) \Bigr)
- K \: \overline{a}(x,t) \: \Bigl( 1 - \overline{a}(x,t) \Bigr)^2,
\end{equation}
\nrtxt
where $D = (m N h)^{-1}$, $K = {N h \over N-1}$. Note that, as defined, 
$K$ is inversely proportional to the diffusion coefficient $D$, $K = \alpha / D$, 
where $\alpha = {1 \over m \: (N-1)}$. 

	Equation (\ref{a_bar_ev_ctime}) can be written in terms of the same 
variables as the mean-field reaction-diffusion equation (\ref{mf_RD_resc}), 
with a rescaling of space, $x \rightarrow x / \sqrt{D}$, as follows:
\wdtxt
\begin{equation}
\label{a_bar_ev_ctime_final}
{\partial \overline{a}(x,t) \over \partial t} = {\partial^2 \overline{a}(x,t)
\over \partial x^2} + \overline{a}(x,t) \: \Bigl( \beta - \overline{a}(x,t) \Bigr)
- {\alpha \over D} \; \overline{a}(x,t) \: \Bigl( 1 - \overline{a}(x,t) \Bigr)^2.
\end{equation}
\nrtxt
The reaction rate in eq. (\ref{a_bar_ev_ctime_final}) is not quadratic 
in $\overline{a}(x,t)$, in contrast to eq. (\ref{mf_RD_resc}), because of the 
$\cal{O}$$(D^{-1})$ correction term. For convenience, all further analysis 
will be carried out using the continuous-space, continuous-time equation
(\ref{a_bar_ev_ctime_final}) rather than the finite-difference equation 
(\ref{a_bar_ev_final}). Since eq. (\ref{a_bar_ev_final}) is obtained from eq. 
(\ref{a_bar_ev_ctime_final}) by Euler discretization and a linear transformation 
of variables, both equations describe essentially the same dynamics.

\subsection{Analysis of the wave front solutions}
	Following the standard formalism \cite{fife,scott,murray}, we recast 
eq. (\ref{a_bar_ev_ctime_final}) in terms of the variable $z$, the coordinate 
of a reference frame moving with the wave front, i.e. $z = x - v t$,
\wdtxt
\begin{equation}
\label{a_bar_ev_z}
{d^2 \overline{a} \over d z^2} + v {d \overline{a} \over d z} = 
\overline{a} \: \left(\overline{a} - \beta \right) + {\alpha \over D} \; 
\overline{a} \: \left( 1 - \overline{a} \right)^2.
\end{equation}
\nrtxt
Here we will consider only wave fronts moving to the right ($v > 0$); generalization 
of our results to the left-moving fronts is trivial. Letting $p = d \overline{a}
/ d z$, we can rewrite eq. (\ref{a_bar_ev_z}) as a system of first-order ODEs,	      
\addtocounter{equation}{1}
\[
\left \{
\begin{array}{rcl}	    
\dot{\overline{a}} & = & p, \\						
 & & \hspace{175 pt} (\theequation) \\
\dot{p} & = & - v p + \overline{a} \left( \overline{a} - \beta \right) +
{\alpha \over D} \: \overline{a} \left( 1 - \overline{a} \right)^2,
\end{array}				       
\right.
\]
where the dot indicates differentiation with respect to $z$. This system posesses
up to three fixed points, one at the origin of the phase plane and the other 
two at $(\overline{a},p) = (\overline{a}_{1s},0) \; {\rm and} \; (\overline{a}_{2s},0)$,
where
\begin{eqnarray}
\label{mobile_fx_pt}
\overline{a}_{1s} &=& 1 - {D - \sqrt{D^2 + 4 \alpha (\beta - 1) D} 
\over 2 \alpha}, \nonumber \\
\overline{a}_{2s} &=& 1 - {D + \sqrt{D^2 + 4 \alpha (\beta - 1) D} 
\over 2 \alpha}. 
\end{eqnarray}

\begin{figure}[htbp]
\begin{center}
\leavevmode
\epsffile{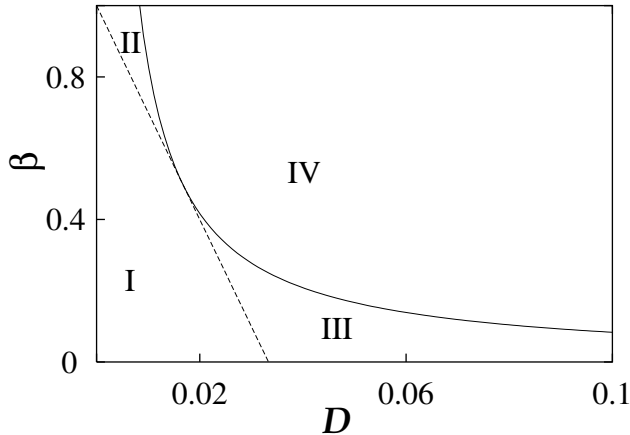}
\end{center}
\caption{Diagram of the fixed-point structure of the ODE system 
(\addtocounter{equation}{-2}\theequation\addtocounter{equation}{2}) as a 
function of $D$ and $\beta$, with $\alpha = {1 \over 120}$. The critical 
diffusion coefficient line $D_{{\rm cr}} = \alpha / \beta$ is indicated
by a solid line. The dashed line denotes the locus of ($D, \beta$) values 
where the movable fixed points first emerge as a degenerate node. For 
description of the fixed-point behaviour in the different zones in the diagram, 
see text above.}
\label{fig1}
\end{figure}

	From a linear stability analysis, we find the following relation for the 
eigenvalues associated with the fixed points:
\begin{equation}
\label{eigenv}
\mu^2 + v \mu - {\alpha \over D} \: \left( 1 - \overline{a} \right) \:
\left( 1 - 3 \overline{a} \right) - 2 \overline{a} + \beta = 0,
\end{equation}
where $\overline{a} = 0, \overline{a}_{1s}, \; {\rm or} \; \overline{a}_{2s}$.
The diagram in Fig. \ref{fig1} summarizes the behaviour of the fixed-point
structure with respect to $D$ and $\beta$, as determined by eqns. (\ref{mobile_fx_pt})
and (\ref{eigenv}). The four zones in the diagram correspond to different 
linear stability properties of the fixed points as follows: 
\newcounter{zones}
\vspace*{-12 pt}
\begin{list}{zone \Roman{zones}:}{\usecounter{zones}}
\item{only the fixed point at the origin exists and is unstable,}
\end{list}

\begin{minipage}[b]{8 cm}
\vspace*{4 pt}
\begin{list}{zone \Roman{zones}:}{\usecounter{zones}}
\addtocounter{zones}{1}
\item{the point at the origin is unstable, the point at $\overline{a}_{1s}$ 
is unstable and moves away from the origin as $D$ increases ($\overline{a}_{1s} > 0$), 
the point at $\overline{a}_{2s}$ is stable and moves toward the origin as $D$
increases ($\overline{a}_{2s} > 0$),}
\vspace*{12 pt}
\item{the point at the origin is unstable, the point at $\overline{a}_{1s}$ 
is stable and moves toward the origin as $D$ increases ($\overline{a}_{1s} < 0$), 
the point at $\overline{a}_{2s}$ is unstable and moves away from the origin 
as $D$ increases ($\overline{a}_{2s} < 0$), and}
\vspace*{12 pt}
\item{the point at the origin is stable, the point at $\overline{a}_{1s}$ 
is unstable and moves away from the origin as $D$ increases ($\overline{a}_{1s} > 0$), 
the point at $\overline{a}_{2s}$ is unstable and moves away from the origin 
as $D$ increases ($\overline{a}_{2s} < 0$).}
\end{list}
\end{minipage}

	The two lines that define these zones are: a) $D = 4 \alpha (1 - \beta)$,
the line of parameter values at which the fixed points at $\overline{a}_{1s}$ 
and $\overline{a}_{2s}$ first emerge as a stable, degenerate, zero-eigenvalue 
node (dashed line), and b) $D = \alpha / \beta$, the line where the fixed point 
at the origin changes its stability (solid line). We observe that formation of 
a wave front replacing the steady state at the origin with the steady state at 
$\overline{a}_{1s}$ is possible if the diffusion coefficient is greater than
the critical value $D_{{\rm cr}} = \alpha / \beta$. Indeed, in that region of 
the diagram the fixed point at the origin is stable, the point at $\overline{a}_{1s}$
is unstable (with $\overline{a}_{1s} > 0$), and $\overline{a}_{2s}$ is negative.
Moreover, it can be easily seen that $\overline{a}_{1s}$ tends to $\beta$ as 
$D$ increases. Thus, in the limit $D \rightarrow \infty$ the fixed-point 
structure of (\addtocounter{equation}{-2}\theequation\addtocounter{equation}{2}) 
is equivalent to that of eq. (\ref{mf_RD_resc}) of Sec. \ref{model}, save for 
additional fixed point in an unphysical area of the phase plane. 

	In order to represent a physically meaningful state of the system, a 
wave front solution of (\addtocounter{equation}{-2}\theequation\addtocounter{equation}{2})
with $ 0 < \beta < 1$ must exhibit no oscillatory behaviour near the 
leading edge of the front. This implies that both roots of eq. (\ref{eigenv}) 
with $\overline{a} = 0$ should be real. One can show by elementary analysis 
that this requirement is satisfied if
\begin{equation}			    
\label{vmin}
v \geq 2 \sqrt{\beta - {\alpha \over D}} \; .
\end{equation}	    
For any finite $D$ the minimum propagation velocity given by 
(\ref{vmin}) and the concentration at the upper plateau of the wave front 
($\overline{a}_{1s}$ in eq. (\ref{mobile_fx_pt})) are both less than their 
respective values predicted by the mean-field theory and converge to those 
values as $D \rightarrow \infty$. 
	
\section{Numerical Results}
\label{LGA}
	In order to test the accuracy of the results obtained in the preceding
section, we numerically solved the full Markov chain dynamics using a lattice-gas 
automaton algorithm. The simulations were performed on a thin ($4 \times 3000$ 
nodes) strip of a 2-dimensional triangular lattice; wave fronts were generated 
from a step-like initial concentration profile placed in the middle of the 
lattice. The value of the rate coefficient $k_2$ used in these simulations was 
$0.333$, so that $\beta = 0.667$. The exclusion parameter was fixed at $N = 21$
which, together with the coordination number for triangular lattice ($m = 6$), 
gives $\alpha = {1 \over 120}$. The value of the diffusion coefficient $D = (m N h)^{-1}$ 
varied between $0.1$ and $1.0$. All of the front properties were measured using 
the mean concentration profile obtained by averaging over $10$ realizations of 
the dynamics. Sufficient time was allowed to elapse before the measurements 
were taken to ensure that the wave front relaxed to its stable, uniformly 
propagating form. In our simulations, the period of transient dynamics was 
found to be relatively long ($\sim 10^5$ automaton time steps). Within this 
period, the front width was observed to grow diffusively; however, no significant 
growth in the front width was detected after the transient regime.

\begin{figure}[htbp]
\begin{center}
\leavevmode
\epsffile{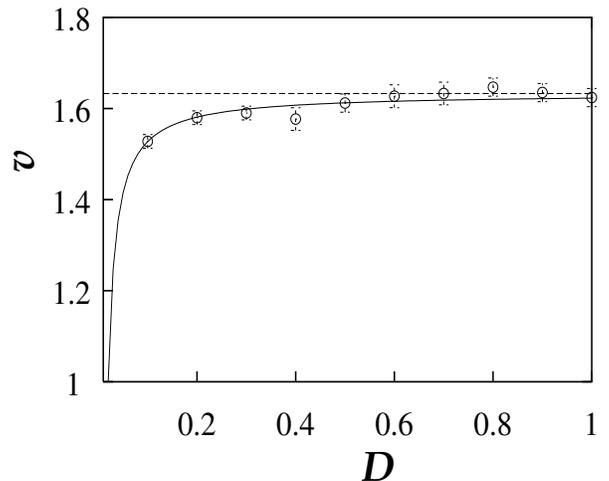}
\end{center}
\caption{Dimensionless front velocity $v$ versus the diffusion coefficient $D$.
Circles denote the data obtained in lattice-gas automaton simulations. The 
minimum propagation velocity given by the present theory (eq. (\ref{vmin})) is
plotted with a solid line. The results of the mean-field theory are indicated 
by a dashed line. Parameters $\alpha$ and $\beta$ are equal to ${1 \over 120}$ 
and $0.667$, respectively.}
\label{fig2}
\end{figure}

	In Fig. \ref{fig2} we compare the front velocity observed in these 
simulations as a function of the diffusion coefficient $D$ with the minimum 
propagation velocity predicted on the basis of the present theory (eq. (\ref{vmin}))
and the mean-field, reaction-diffusion equation (\ref{mf_RD_resc}). 
We observe that the propagation velocity predicted by the present theory is in 
good agreement with the data from the numerical simulations.

\section{Conclusions}
\label{conclusions}
	The generalized kinetic equation derived in this study admits travelling 
wave solutions whose properties are different from the predictions of mean-field 
theory. There is a non-zero critical value of the diffusion coefficient below 
which no stable wave front solution exists. Moreover, the concentration at the 
upper plateau of the wave front varies with the diffusion coefficient as well 
as the kinetic parameters of the system. Most significantly, the (dimensionless) 
minimum front velocity decreases as the diffusion coefficient is decreased. 
These effects have their origin in the local correlations built by the fluctuating 
reactive dynamics which persist for finite values of the diffusion coefficient.

	For finite values of the diffusion coefficient, measurements of the front 
velocity in two-dimesional lattice-gas simulations of the full Markov chain 
model confirmed that the present theory describes the departure from the 
mean-field results with quantitative accuracy. Similar simulations were also 
carried out on one-dimensional systems. The results of these simulations showed 
the same qualitative trend as in two dimensions. However, because the primary 
assumption of our theory (the factorization of the full probability distribution) 
is not valid for one spatial dimension, the quantitative agreement between 
theoretical data and the simulation results is much poorer than that for two 
dimensions.

	There is a number of numerical studies \cite{lemarchand,lesne,karzazi,doering} 
on the breakdown of mean-field descriptions for quadratic autocatalysis fronts 
using a variety of microscopic models, none of which is the same as that employed 
in our study.  The lattice-gas automaton simulations of Lemarchand {\it et al.} 
\cite{lemarchand} have shown that the mean-field results remain valid for the 
range of diffusion coefficient values used in their study. Later investigations 
using a Langevin equation approach reported an increase in the front velocity 
above the mean-field value \cite{lesne,karzazi}. Monte Carlo studies of Riordan 
{\it et al.} have focused on the front width and have found that the width 
grows algebraically in one and two space dimensions, while in higher dimensions 
the width dynamics is adequately described by mean-field theory \cite{doering}. 
In our simulations on two-dimensional systems with small transverse size, the 
front width grows diffusively during a very long transient period but then 
saturates and the front evolves without change in width. The effects of microscopic 
fluctuations on quadratic autocatalyis fronts clearly present a number of 
interesting features.

\section*{Acknowledgments}
	This work was partially supported by the Natural Sciences and Engineering 
Research Council of Canada. M.V. also benefits from a Connaught Scholarship.
Computing resources for this study were provided in part by the University of 
Toronto Information Commons.

\section*{Appendix}
	To compute the $\cal{O}$$(\gamma^2)$ correction term in eq. (\ref{a_bar_ev}), 
note the following properties of operator $\hat{W}^D$ (the proof can be obtained 
by straightforward calculation, as in Appendix A of Ref.~\cite{theory}):
\wdtxt
\begin{eqnarray}
\label{WD_prop_1}
\sum_{a({\bf r}) = 0}^{N} a({\bf r}) \left( \hat{W}^D \right)^p X(a({\bf r}),n) &=&
\left(-{1 \over N} \right)^p \: \sum_{a({\bf r}) = 0}^{N} a({\bf r}) X(a({\bf r}),n),
\nonumber \\
\sum_{a({\bf r}) = 0}^{N} a^2({\bf r}) \left( \hat{W}^D \right)^p X(a({\bf r}),n) &=&
\left(-{2 \over N} \right)^p \: \sum_{a({\bf r}) = 0}^{N} a^2({\bf r}) X(a({\bf r}),n),
\end{eqnarray}
\nrtxt
where $\left( \hat{W}^D \right)^p$ is to be understood as a time-ordered 
product of operators $\hat{W}^D$ taken at any $p$ moments of time between $0$ 
and $n$. $X(a({\bf r}),n)$ is any function of $a({\bf r})$ and time which has 
the following properties:
\begin{eqnarray}
\label{averages}
& & \sum_{a({\bf r}) = 0}^{N} X(a({\bf r}), n) = 0, \nonumber \\
& & \sum_{a({\bf r}) = 0}^{N} a({\bf r}) \: X(a({\bf r}), n) = 0.
\end{eqnarray}

       We remove one of the operators $\hat{W}^D$ by adding the inverse 
operator $\left( \hat{W}^{D} \right)^{-1}$, taken at an appropriate moment of 
time, to the left of $\left( \hat{W}^D \right)^p$ in (\ref{WD_prop_1}). Bearing 
in mind the property (\ref{WD_prop_1}), we find
\wdtxt
\begin{eqnarray}
\label{WD_prop_2}
\sum_{a({\bf r}) = 0}^{N} a({\bf r}) \: \hat{W}^{D^{-1}} \: \left( \hat{W}^D \right)^p X(a({\bf r}),n) &=&
\left(-{1 \over N} \right)^{p-1} \: \sum_{a({\bf r}) = 0}^{N} a({\bf r}) X(a({\bf r}),n),
\nonumber \\
\sum_{a({\bf r}) = 0}^{N} a^2({\bf r}) \: \hat{W}^{D^{-1}} \: \left( \hat{W}^D \right)^p X(a({\bf r}),n) &=&
\left(-{2 \over N} \right)^{p-1} \: \sum_{a({\bf r}) = 0}^{N} a^2({\bf r}) 
X(a({\bf r}),n).
\end{eqnarray}
\nrtxt
	Note that if (\ref{averages}) holds for $X(a({\bf r}),n)$, it also 
holds for $\left( \hat{W}^D \right)^p X(a({\bf r}),n)$, i.e. $\hat{W}^D$ is a 
function-to-function map within the class of functions satisfying (\ref{averages}).
Furthermore, $\hat{S} P_B(a({\bf r}),n)$ belongs to that class of functions, 
as can be easily seen from the definition of the operator $\hat{S}$ (\ref{S_def}) 
and the following two identities:
\begin{eqnarray}
\label{aux_ident}
& & \sum_{a({\bf r}) = 0}^{N} a({\bf r}) \: \hat{W}^R P_B(a({\bf r}),n) =
R(\overline{a}({\bf r},n)), \nonumber \\
& & \sum_{a({\bf r}) = 0}^{N} a({\bf r}) \: P_B(a({\bf r}),n) =
\overline{a}({\bf r},n).
\end{eqnarray}
Using these observations, we can infer by simple inspection of (\ref{WD_prop_1}) 
and (\ref{WD_prop_2}) that
\wdtxt
\begin{eqnarray}
\label{WD_inv_prop}
\sum_{a({\bf r}) = 0}^{N} a({\bf r}) \: \left( \hat{W}^{D} \right)^{-1} 
\hat{S} P_B(a({\bf r}),n) \: &=& - N \: \sum_{a({\bf r}) = 0}^{N} a({\bf r}) 
\hat{S} P_B(a({\bf r}),n) \nonumber \\ 
&=& 0 \: , \\ \nonumber \\
\sum_{a({\bf r}) = 0}^{N} a^2({\bf r}) \: \left( \hat{W}^{D} \right)^{-1} \: 
\hat{S} P_B(a({\bf r}),n) &=& - {N \over 2} \: \sum_{a({\bf r}) = 0}^{N} 
a^2({\bf r}) \hat{S} P_B(a({\bf r}),n). \nonumber 
\end{eqnarray}
\nrtxt
\hspace{\parindent}
Using (\ref{WReact}) and (\ref{WD_inv_prop}), after a few simple 
transformations, we obtain 
\wdtxt
\begin{eqnarray}
\label{app_corr_term}    
\sum_{a({\bf r}) = 0}^{N} a({\bf r}) \: \hat{W}^R \left( \hat{W}^{D} \right)^{-1} 
\hat{S} P_B(a({\bf r}),n) &=& 
h \sum_{a({\bf r}) = 0}^{N} \left[ \left( {N \over N-1} - k_2 \right) 
\: a({\bf r}) \: - \: {h \over N-1} \: a^2({\bf r}) \right] \: \left( 
\hat{W}^{D} \right)^{-1} \hat{S} P_B(a({\bf r}),n) \nonumber \\
&=& {N h \over 2 (N-1)} \: \sum_{a({\bf r}) = 0}^{N} a^2({\bf r}) \: 
\hat{S} P_B(a({\bf r}),n) \\
&=& {N h^2 \over N-1} \; \overline{a}({\bf r},n) \left(
1 - {\overline{a}({\bf r},n) \over N} \right)^2 \: , \nonumber
\end{eqnarray}
\nrtxt
for the $\cal{O}$$(\gamma^2)$ correction term.

\end{multicols}
\end{document}